\documentclass[twocolumn]{revtex4}
\usepackage{amsfonts,amsmath,amssymb,latexsym,epsfig,color,hyperref}

\hypersetup{
    colorlinks,%
    citecolor=blue,%
    filecolor=blue,%
    linkcolor=blue,%
    urlcolor=blue
}
\definecolor{bluecolor}{rgb}{0,0.,1.}

\definecolor{redcolor}{rgb}{.7,0.,0.}

\newcommand{\pr}[1]{\left( #1\right)}

\newcommand{\es}[1]{\begin{equation}\begin{split}#1\end{split}\end{equation}}

\newcommand{\R}{\mathbb{R}}
\newcommand{\N}{\mathbb{N}}

\newcommand{\A}{\mathcal{A}}
\newcommand{\B}{\mathcal{B}}
\newcommand{\NA}{N_{\mathcal{A}}}
\newcommand{\NB}{N_{\mathcal{B}}}
\newcommand{\erf}{\mathrm{erf}}

\begin{document}

\title{Network connectivity through small openings}
\author{Orestis Georgiou$^{1,2}$, Carl P. Dettmann$^{2}$, Justin P. Coon$^{1,3}$}

\affiliation{1 Toshiba Telecommunications Research Laboratory, 32 Queens Square, Bristol, BS1 4ND, UK.}
\affiliation{2 School of Mathematics, University of Bristol, University Walk, Bristol BS8 1TW, UK.}
\affiliation{3 Department of Electrical and Electronic Engineering, University of Bristol, BS8 1UB, Bristol, UK.}

%%%%%%%%%%%%%%%%%%%%%%

\begin{abstract}
Network connectivity is usually addressed for convex domains where a direct line of sight exists between any two transmitting/receiving nodes. Here, we develop a general theory for the network connectivity properties across a small opening, rendering the domain essentially non-convex. 
Our analytic approach can go only so far as we encounter what is referred to in statistical physics as quenched disorder making the problem non-trivial. We confirm our theory through computer simulations, obtain leading order approximations and discuss possible extensions and applications.
\end{abstract}

\maketitle
%%%%%%%%%%%%%%%%%%%%%

%\setlength{\abovedisplayskip}{5pt}
%\setlength{\belowdisplayskip}{5pt}

\section{Introduction}
\label{sec:intro}

Wireless ad hoc and sensor networks \cite{cordeiro2011ad} have attracted much attention in recent years as they do not rely on a pre-existing system infrastructure, such as central routers or access points. 
Instead communication devices, here referred to as nodes, can relay messages to one another in a multi-hop fashion, thus achieving good coverage and connectivity over large areas. 
Modern applications of wireless ad hoc and sensor mesh networks can be found \textit{inter alia}, in swarm robotics \cite{mohan2009extensive} deployed in hazardous locations such as factories, mines and disaster areas \cite{li2009connectivity}, laptops,
power distribution (``smart grid") technologies \cite{moslehi2010reliability}, vehicles for road safety \cite{harri2009mobility}, and underwater sensor networks \cite{heidemann2012underwater}.

Achieving smooth functionality of a network in the form of good connectivity, whilst simultaneously minimizing economic, computational and energy costs is a difficult task, addressed from a physical layer's perspective by the theory of random graphs \cite{penrose2003random} and complex networks  \cite{albert2002statistical}. 
Here, networks are modelled by a collection of nodes randomly distributed in a region of space, pairwise connected with a relative position dependent probability.
The network's connectivity properties can then be characterized via a plethora of methods and measures such as for example various clustering statistics, network modularity measures, node importance and correlations between degrees of neighbouring nodes. 
Of particular interest in dense networks is the optimal number of nodes necessary to maintain full connectivity with certain probability \cite{xue2004number}.

A theory for predicting the probability $P_{fc}$ with which a random network is fully connected in confined convex spaces was recently developed in \cite{coon2012impact} and \cite{coon2012full}. 
There, it was shown that $P_{fc}$ is strongly influenced by the details of the domain boundary and in fact is asymptotically dominated by the lack of connectivity near the sharpest interior corner.
The main result was a general analytic formula facilitating the design and analysis of reliable wireless mesh networks \cite{haenggi2009stochastic}.

\begin{figure}[t]
\begin{center}
\includegraphics[scale=0.33]{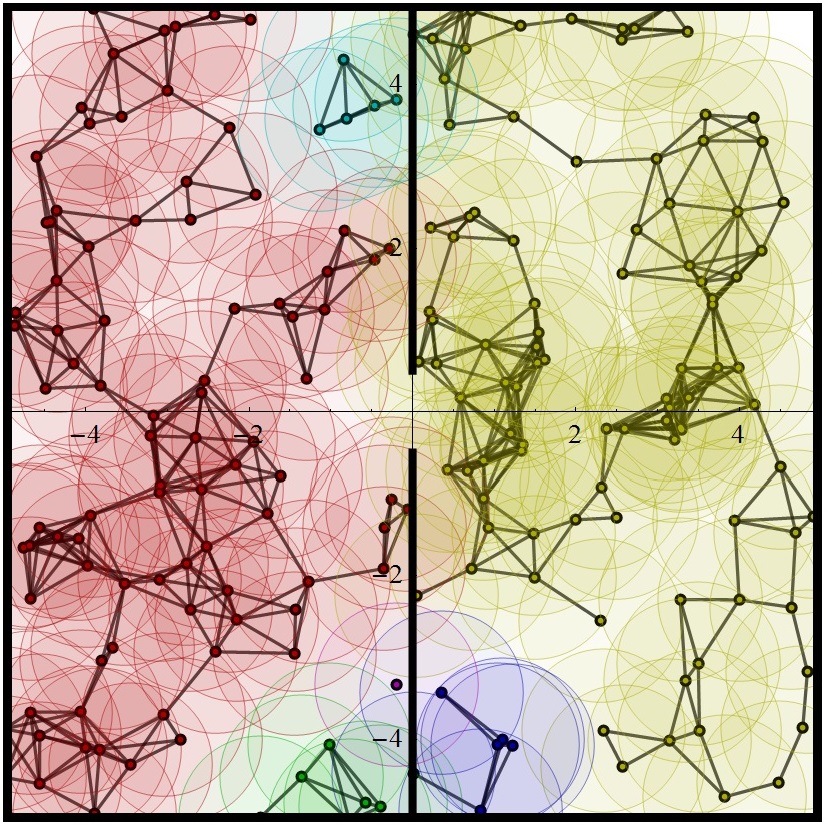}
\caption{\label{fig:setup} Clustering of a random network  formed in a square domain of side $L=10$ at density $\rho=2$. A hard vertical wall with a keyhole opening of width $w=1$ splits the domain into two equal parts thus hindering full connectivity.}
\end{center}
\end{figure}

In this paper, we extend this theory to non-convex domains and in particular to domains involving a small \textit{keyhole} type opening as shown in Fig.~\ref{fig:setup}.
The notion of the keyhole problem in wireless communications was introduced in \cite{keyholeorigin}, and has since been extensively studied from the antenna propagation perspective and the relevant channel distributions \cite{nrayleigh}.
Instead, here we focus on the network aspect of the keyhole problem and explain why this system exhibits semi-quenched disorder, a feature common in spin glass systems (e.g. disordered magnets with frustrated interactions) \cite{sornette2004critical}. 
As our main contribution, we derive a general formula for the probability of two sub-networks to connect through a small opening.
We confirm our results using computer simulations and furthermore obtain leading order asymptotic expansions in closed form.
To the best of our knowledge, together with the recent works of \cite{bocus2012keyhole}, our results constitute the first analytic approach to network connectivity in non-convex domains.

The paper is structured as follows: 
In Sec.~\ref{sec:setup} we describe the keyhole problem set-up and define the relevant parameters and observables. 
In Sec.~\ref{sec:sub} we recall how the probability of full connectivity is calculated in convex domains and in Sec.~\ref{sec:full} we reduce the problem of full network connectivity to that of the existence probability of a bridging link between sub-domains.
In Sec.~\ref{sec:lead} we obtain leading order analytic approximations to this probability and in Sec.~\ref{sec:quench} discuss why the system exhibits quenched disorder and thus derive our main result, a general formula for the probability of two sub-networks to connect through a small opening.
In Sec.~\ref{sec:num} we confirm our results through computer simulations and discuss asymptotic limitations and difficulties due to the many length scales associated with the problem at hand.
Finally, some concluding remarks are given in Sec.~\ref{sec:concl}.

%%%%%%%%%%%%%%%%%%%%%%%%%

\section{Description of the Problem}
\label{sec:setup}

In its simplest form, the problem set-up involves a domain $\mathcal{V}\subset \R^2$ of area $V$ which is split into two convex sub-domains $\A,\B\subset\mathcal{V}$, containing $\NA$ and $\NB$ nodes respectively.
The two domains are separated by a hard straight wall containing a small \textit{keyhole} opening of width $w$ (see Fig.~\ref{fig:setup}). 
For reasons to be discussed later, we assume that the keyhole is located sufficiently away from the remaining borders of $\A$ and $\B$.
The node locations $\textbf{a}_i \in \A$ and $\textbf{b}_j \in \B$ for $i\in[1, \NA]$ and  $j\in[1, \NB]$ are chosen at random from a uniform distribution with density $\rho_\mathcal{A,B}= N_{\mathcal{A,B}}/V_{\mathcal{A,B}}$ where $V_{\mathcal{A,B}}$ is the corresponding volume of the sub-domains.

After deployment of the nodes, communication links between pairs of nodes are established with probability $H(r)$, where $r$ is the relative distance between the pair.
We maintain physical relevance by adopting a specific pair connectedness function $H(r)$ derived from wireless communication theory~\cite{tse2005fundamentals} and applicable to ad hoc and sensor networks - the Rayleigh fading model \footnote{The Rayleigh fading link model assumes that the received signal is a superposition of a large number of randomly phased waves (a result of small scale scattering effects).}.
Here, the information outage probability $P_{out}$ due to a single-input single-output(SISO) \footnote{For multiple input multiple output (MIMO) systems the total channel gain through a keyhole is modelled by products of random variables \cite{nrayleigh}.} system is given by $P_{out}=\textrm{Pr}\pr{\log_2 (1+ \textrm{SNR} \times |h|^2) < x}$, where $x$ is the minimum outage rate threshold, $\textrm{SNR} \propto r^{-\eta}$ is the signal to noise ratio and $\eta$ is the path loss exponent \footnote{Typically $\eta=2$ corresponds to propagation in free space but for cluttered environments it is observed to be $\eta>2$.}.
Finally, $h$ is the channel transfer coefficient and $|h|^2$ is modelled as an exponentially distributed random variable.
Hence, the connection probability between two nodes a distance $r$ apart can be expressed as
\es{
H(r)=1- P_{out}= e^{-\beta r^{\eta}}
\label{H},}
where $\beta$ sets the characteristic connection length $r_0=\beta^{-1/\eta}$.
It is worth noting that in the limit of $\eta\rightarrow \infty$, the connection between nodes is no longer probabilistic and converges to the well studied case in geometric graph theory~\cite{penrose2003random}, the \textit{unit disk} model with an \textit{on/off} connection range at the limiting $r_0$.

Figure~\ref{fig:setup} shows a typical realization of a random network in a square domain of side $L=10$ with $w=1$ at a density of $\rho_\A=\rho_\B=2$ using $\beta=1$ and $\eta=\infty$.
We observe that two large connected components (clusters) are formed, one in each sub-domain, with smaller ones forming near some of the corners.
In this paper, we will investigate the probability of achieving a fully connected network denoted by $P_{fc}^{(\mathcal{V})}$ for a given pair of densities $\rho_\A$ and $\rho_\B$.

%%%%%%%%%%%%%%%%%%%%%%

\section{Full Connectivity in Sub-Domains}
\label{sec:sub}

In \cite{coon2012full}, it was shown that the probability of a spatially confined random network being fully connected $P_{fc}$ is dominated by local boundary effects due to ``hard to connect to'' regions. 
Thus, contrary to the expected universal features of large networks, full connectivity is governed by local and not global features of the network domain.
Through the use of a cluster expansion, often used in statistical physics, it was shown that for arbitrary convex domain $\mathcal{V}\subset \R^{d}$
\es{P_{fc} = 1-\rho \int_{\mathcal{V}} e^{-\rho\int_{\mathcal{V}} H(r_{12}) \mathrm{ d}\mathbf{r}_{1} } \mathrm{ d}\mathbf{r}_{2}
,
\label{pfc}}
where $r_{12}= |\mathbf{r}_1 - \mathbf{r}_2|$ is the relative distance between nodes with position coordinates $\mathbf{r}_1 , \mathbf{r}_2 \in \mathcal{V}$. 
For fast decaying connectivity functions $H(r)$ (as in Eq.~\eqref{H}), the outer integral in \eqref{pfc} can be approximated by a sum over separate boundary objects of different co-dimension $i=0,1,\ldots d$ (with $i=0$ corresponding to the bulk/volume term)
\es{\label{gen1}P_{fc}  \approx  1- \sum_{i=0}^{d}\sum_{j_{i}} \rho^{1-i} \mathbf{G}_{j_{i}} \mathbf{V}_{j_{i}} e^{-\rho \omega_{j_{i}} \int_{0}^{\infty}r^{d-1}H(r)\mathrm{d}r }
,}
where $d\in\N$ is the space dimension, $\mathbf{G}_{j_{i}}$ is a geometrical factor for each object $j$ of co-dimension $i$ while $\mathbf{V}_{j_{i}}$ is the corresponding $d-i$ dimensional volume of the boundary object with internal solid angle $\omega_{j_{i}}\in(0,2\pi^{d/2}/ \Gamma(d/2) )$.

Equation \eqref{gen1}, suggests that to a good approximation network connectivity is the sum of its parts.
For example, using $\eta=2$ in \eqref{H} we have that a random network of $N$ nodes confined in a convex $n$-gon of area $\mathbf{V}_1$, perimeter $\mathbf{V}_2$ and interior angles $\Omega=\{ \omega_1,\omega_2 \ldots \omega_n \}$, will be fully connected with probability
\es{
P_{fc} = 1- \mathbf{V}_1 \rho e^{-\frac{\pi}{\beta} \rho } - \mathbf{V}_2 \sqrt{\frac{\beta}{\pi}}  e^{-\frac{\pi}{2\beta} \rho} - \sum_{i=1}^{n}\frac{4\beta  e^{-\frac{\omega_i}{2\beta} \rho}}{\rho\pi \sin(\omega_i)}  
,
\label{sub} }
where $\rho=N/\mathbf{V}_1$.
Therefore, for the keyhole set-up of Sec.~\ref{sec:setup}, it follows that full connectivity in each sub-domain $P_{fc}^{(\A)}$ and $P_{fc}^{(\B)}$ is given by \eqref{sub}. 
Notice that at high densities $\rho$, the term with the sharpest corner $\min (\Omega)$ will dominate $P_{fc}$.

%%%%%%%%%%%%%%%%%%%%%%%%%%

\section{Full Connectivity}
\label{sec:full}

\begin{figure}[t]
\begin{center}
\includegraphics[scale=0.31]{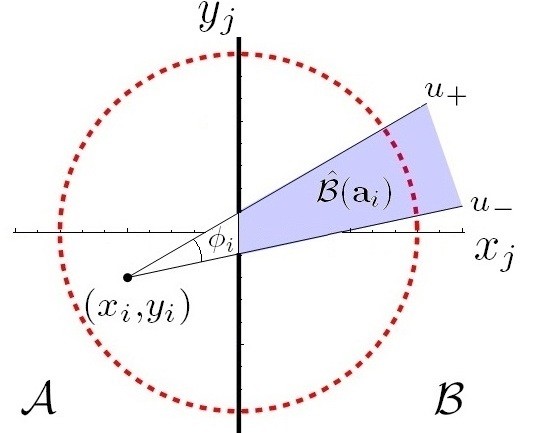}
\caption{\label{fig:near} Close-up of the keyhole opening showing a node $\textbf{a}_{i}$ in sub-domain $\A$ and its associated connection region  $\hat{\B} (\textbf{a}_{i})\subset{\B}$. The dashed circle simply illustrates that the keyhole is sufficiently isolated from any other boundary elements of the domain $\mathcal{V}$.}
\end{center}
\end{figure}

If both sub-domains are fully connected, and at least one ``bridging'' link exists through the keyhole opening, then the combined network is also fully connected.
Thus we may write
\es{
P_{fc}^{(\mathcal{V})}=P_{fc}^{(\A)} P_{fc}^{(\B)} X
,\label{inde}}
where $X$ is the probability of at least one bridging link between the two sub-domains.
As with the simplification of \eqref{pfc} into \eqref{gen1}, the assumption of statistical independence between the three events in \eqref{inde} is justified due to the fast decaying $H(r)$.
For $w$ sufficiently small, no node in $\A$ is likely to influence the connectivity properties of nodes in $\B$, and \textit{vice versa}.
Writing $P_{fc}^{(\mathcal{V})}$ as in \eqref{inde} is particularly helpful when performing computer simulations as calculating $X$ is typically much faster than calculating $P_{fc}^{(\mathcal{V})}$ (see Sec.~\ref{sec:num}).

Noting that the probability of at least one bridging link is just the complement of no bridging links, we have that
\es{
X= 1- \langle \langle \prod_{i=1}^{\NA} \prod_{j=1}^{\NB}
(1- \chi_{ij}  H_{ij}) \rangle_{\B}  \rangle_{\A},
\label{X}
}
where $H_{ij}= H(|\textbf{a}_{i} - \textbf{b}_{j}|)$ and $\chi_{ij}$ equals $1$ if a ``line of sight'' exists between nodes $\textbf{a}_{i}$ and $\textbf{b}_{j}$, and is zero otherwise.
The angled brackets in \eqref{X} represent a spatial average over all possible node configurations within each sub-domain
\es{\langle O\rangle_{\mathcal{C}} = \frac{1}{V_{\mathcal{C}}^{N_{\mathcal{C}}}}\int_{\mathcal{C}^{N_{\mathcal{C}}}} O(\mathbf{c}_{1},\mathbf{c}_{2},\ldots, \mathbf{c}_{N_{\mathcal{C}}} ) \mathrm{ d}\mathbf{c}_{1} \mathrm{ d}\mathbf{c}_{2}\ldots \mathrm{ d}\mathbf{c}_{N_{\mathcal{C}}}
,}
with $\mathcal{C}=\A,\B$, and $\textbf{c}=\textbf{a}, \textbf{b},$ respectively.
Note that the order in which the average in \eqref{X} is performed is not important.

%%%%%%%%%%%%%%%%

\section{Leading order approximation}
\label{sec:lead}

The crudest approximation of $X$ can be obtained by assuming that all spatial integrals in \eqref{X} separate such that
\es{
X &\approx  1- \pr{1-  \langle \langle \chi_{ij} H_{ij}\rangle_{\B} \rangle_{\A}  }^{\NA \NB  } \\
&= 1- \textrm{exp} \pr{ -\rho_{\A} \rho_{\B} \int_{\A} \int_{\B}  \chi_{ij}H_{ij} \textrm{d}\textbf{b}_{j}  \textrm{d}\textbf{a}_{i}} 
,
\label{X1}}
where we have assumed that $\NA,\NB \gg 1$ and have used the fact that $(1-x)^N =e^{-N x}$ in the limit of $N\rightarrow \infty$. 
We shall be using this approximation throughout our analysis.

In Eq.~\eqref{X1} we can identify $\langle \langle \chi_{ij} H_{ij}\rangle_{\B} \rangle_{\A}$ as the probability of a randomly selected pair of nodes to form a bridging link. 
We now calculate the integrals in \eqref{X1} for a \textit{soft} connectivity function $H(r)$ with $\eta=2$.
In doing so $H(r)$ is a Gaussian function thus rendering the mathematics tractable.
For every node $\textbf{a}_i= (x_{i},y_{i})$, there is a line of sight connection region $\hat{\B}(\textbf{a}_i) \subset \B$ (see Fig.~\ref{fig:near}). Hence we have that $\int_{\B}  \chi_{ij}H_{ij} \textrm{d}\textbf{b}_{j} = \int_{\hat{\B}(\textbf{a}_i)} \!\!  H_{ij} \textrm{d}\textbf{b}_{j}$.
In Cartesian coordinates, we get
\es{
\int_{\hat{\B}(x_i ,y_i)} \!\!\!\!\!\! H_{ij} \textrm{d}\textbf{b}_{j}
&= \int_{0}^{\infty} \int_{u_{-}}^{u_{+}}   e^{-\beta \pr{(x_i - x_j)^2 +(y_i - y_j)^2 }} \textrm{d} y_{j} \textrm{d} x_{j} \\
&= \frac{\sqrt{\pi }}{2\sqrt{\beta}}\int_{0}^{\infty} e^{-\beta(x_i-x_j)^{2}} \Big(  \erf(\sqrt{\beta} (u_{+} - y_i) ) \\
&+ \erf(\sqrt{\beta} (y_i - u_{-}) ) \Big) \textrm{d} x_{j},
%&=\int_{0}^{\infty} f(x_{j},x_{i},y_{i}) \textrm{d} x_{j},
\label{int}}
where $u_{\pm}=(y_i \mp w/2) x_{j}/x_i \pm w/2$ are as seen in Fig.~\ref{fig:near}. 
Setting the upper integration limit of $\textrm{d} x_j$ to infinity in an otherwise finite domain $\mathcal{V}$, is justified by the exponentially decreasing integrand function.

The change of variables $(\hat{x}_i,\hat{y}_i,\hat{x}_j )= 1/\sqrt{\beta} (x_i,y_i,x_j)$ simplifies the expression in \eqref{int} whilst a further change of variable $\hat{z}_j=-\hat{x}_j / \hat{x}_i$ eliminates the $\hat{x}_i$ dependence in the argument of the error functions. 
We may now change the order of integration and perform the $\hat{x}_i$ integral over the negative real line first to obtain
\es{
\frac{\erf\big[(1+\hat{z}_{j})(\hat{y}_i+\frac{\sqrt{\beta}w}{2}) \big] - \erf\big[(1+\hat{z}_{j})(\hat{y}_i - \frac{\sqrt{\beta}w}{2}) \big]}{2(1+\hat{z}_j)^2}
,}
which is an even function in $\hat{y}_{i}$. 
Integration in $\hat{y}_{i}\in(-\infty,\infty)$ gives $\sqrt{\pi}w/( 2(1+\hat{z}_j)^2\beta^{3/2} )$ and finally integration in $\hat{z}_{j}\in(0,\infty)$ leads to the immense simplification of
\es{
\int_{\A} \int_{\B}  \chi_{ij}H_{ij} \textrm{d}\textbf{b}_{j}  \textrm{d}\textbf{a}_{i} = \frac{\sqrt{\pi} w}{2\beta^{3/2}},
\label{simp}}
and hence that 
\es{
X\approx 1- \exp\pr{-\rho_{\A} \rho_{\B} \frac{\sqrt{\pi} w }{ 2 \beta^{3/2}} }
\label{XX}.
}
Indeed, Eq.~\eqref{XX} offers a decent approximation of $X$ at low densities and for $w\ll 1$. 
This is confirmed in Fig.~\ref{fig:numerics} where we compare against computer simulations of $X$.
We postpone discussing the accuracy of \eqref{XX} for Sec.~\ref{sec:num}.

\begin{figure*}[t]
\begin{center}
\includegraphics[scale=0.235]{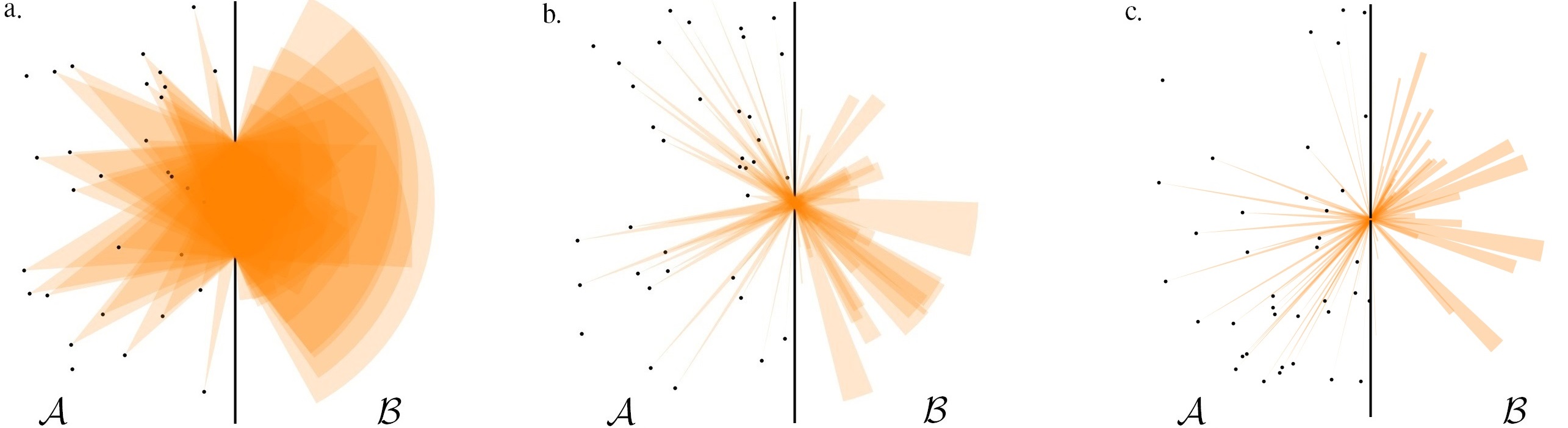}
\caption{\label{fig:overlaps} Overlapping sectors of radius $r_{0}=1$ due to nodes in $\A$. 
The union of these sectors intersected with $\A$ defines the cumulative connection region $\bigcup_i \hat{\B}(\textbf{a}_i)$. 
Note that some nodes are too far away from the keyhole opening and so do not contribute to $\bigcup_i \hat{\B}(\textbf{a}_i)$.
For larger openings (left) the cumulative connection region is dominated by the nearest node to the keyhole, while for smaller openings (middle) the cumulative connection region defines a complicated region. For very small openings (right), there is very little overlap between sectors.}
\end{center}
\end{figure*}

For a \textit{hard} connectivity function $H(r)$ with $\eta=\infty$ and connectivity range $r_0$, the sector angle of $\hat{\B}(\textbf{a}_i)$ is given in polar coordinates by $\phi_i = -w \cos\theta_i / r_i +\mathcal{O}(w^2)$ (see Fig.~\ref{fig:near}). 
Hence we have that 
\es{
\int_{\B} \chi_{ij}H_{ij} \textrm{d} \mathbf{b}_j \approx \frac{\phi_i}{2\pi} \pr{\pi r_0^2 - \pi r_i^2}
,\label{area}}
where we have approximated $\hat{\B}(\textbf{a}_i)$ by an annular sector with radii $r_i<r_0$.
Integrating over $\A$ we have that
\es{
\int_{\pi /2}^{3\pi /2} \int_{0}^{r_0} r_i \frac{\phi_i}{2}(r_0^2 - r_i^2) \textrm{d}r_i \textrm{d}\theta_i
= \frac{2 r_0^3 w}{3}.
\label{hard}}
Notice that substituting $r_0=\beta^{-2}$ into \eqref{hard} gives the correct scaling as in \eqref{simp}. The pre-factor however $2/3 < \sqrt{\pi}/2\approx 0.886$ suggests that connectivity through small openings is better for soft connectivity functions $H(r)$.

%%%%%%%%%%%%%%%%%%%%%%%%%%%%%

\section{Semi-Quenched Disorder}
\label{sec:quench}

Equation \eqref{X1} suffers from the unphysical assumption of uncorrelated node positions. That is, if one node in $\B$ is situated very close to the keyhole opening, then almost all nodes in $\A$ will have an increased probability of connecting to it and \textit{vice versa}. 
Thus, Eq.~\eqref{X1} fails to capture this feature of the network as it assumes that the node positions of the selected pair under investigation are independently chosen at each check. 
In statistical physics, this situation resembles that of \textit{annealed} disorder where random variables are allowed to evolve with time.

The opposite of annealed disorder is 
\textit{quenched} disorder. Here, random variables do not evolve with time, i.e. they are quenched or \textit{frozen}. 
This can be understood best for hard connectivity functions $H(r)$ i.e. when $\eta=\infty$. 
For a particular realization of the $\NA$ node positions in sub-domain $\A$, the union of all connection regions $\bigcup_i  \hat{\B}(\textbf{a}_i)$ forms a cumulative connection region
and grows with $\NA$
% and is bounded from above by $\pi/2 + w$ 
(see Fig.~\ref{fig:overlaps}). 
Since nodes in $\B$ are uniformly distributed, the probability of no bridging links is equal to $(1-|\bigcup_i \hat{\B}(\textbf{a}_i)|/V_{\B})^{\NB}$, and so on average 
\es{
X&= 1-  \pr{1- \frac{\langle |\bigcup_i \hat{\B}(\textbf{a}_i)| \rangle_{\A}}{V_{\B}}  }^{\NB} 
%&= 1- e^{- \rho_{\B} \langle |\bigcup_i \hat{\B}(\textbf{a}_i)| \rangle_{\A}  }
\label{x3}
,}
where $\langle |\bigcup_i \hat{\B}(\textbf{a}_i)| \rangle_{\A} $ is the average area of the cumulative connection region due to all possible spatial configurations of nodes in $\A$.

For small openings relative to the typical node connection range $r_0$, the cumulative connection region $\bigcup_i \hat{\B}(\textbf{a}_i)$ is a complicated object (see Fig.~\ref{fig:overlaps}.b.) leaving little hope in obtaining an expression for its expected area in closed analytic form.
Nevertheless, Eq.~\eqref{x3} provides further insight into the problem as it suggests that only nodes in one of the two sub-domains are quenched; the system is \textit{semi}-quenched.
Therefore we may simplify \eqref{X} to obtain our main result
\es{
X&= 1- \langle \prod_{i=1}^{\NA} \langle \prod_{j=1}^{\NB}
(1- \chi_{ij}  H_{ij}) \rangle_{\B}  \rangle_{\A} \\
%&=1- \langle \prod_{i=1}^{\NA}\pr{1-\langle \chi_{ij}H_{ij}  \rangle_{\B}}^{\NB} \rangle_{\A} \\
&=1- \pr{\langle \big( 1-\langle \chi_{ij}H_{ij}  \rangle_{\B} \big)^{\NB} \rangle_{\A}}^{\NA} \\
%&= 1- \pr{\langle e^{-\NB \langle \chi_{ij}H_{ij}  \rangle_{\B}}
%\pr{ 1- \mathcal{O}(1/V_{\B})}
% \frac{\NB \langle \chi_{ij}H_{ij}  \rangle_{\B}^2}{2} +\ldots} 
%\rangle_{\A}}^{\NA}\\
&= 1- \pr{ \frac{1}{V_\A}\int_{\A} e^{-\rho_{\B} \int_{\B} \chi_{ij}H_{ij} \textrm{d}\textbf{b}_{j}} \textrm{d}\textbf{a}_{i}}^{\NA}
\label{X3}
,}
where we recognize the term in the final brackets as the probability of node $i$ not connecting with any of the $\NB$ nodes in $\B$.
Note that Eq.~\eqref{X3} is symmetric under $\A \leftrightarrow \B$  and is also valid for higher dimensional domains.

%%%%%%%%%%%%%%%%%%%%%%%

\section{Computer Simulations and Discussion}
\label{sec:num}

In this section we verify Eq.~\eqref{X3} through direct numerical integration using standard routines and comparing with computer simulations. 
The computer code written in C++ uses Monte Carlo methods to estimate $X$ for different densities. At each run, an array of $\NB$ randomly chosen coordinates in $\B$ is initiated. 
A random point is then generated in $\A$, and is ``checked" against all points in the array.
The check is successful if the point in $\A$ has a line of sight \textit{and} a connection link according to $H(r)$ with at least one of the points in $\B$. 
This is repeated up to $\NA$ times, or until a successful check. 
Thus, the worst case computational cost is $\mathcal{O}(\NA \NB)$.
Note that an initial sorting of the array in ascending node-to-opening distance order, would typically offer a significant overall speed up due to the increased likelihood of a successful check.

\begin{figure}[t]
\begin{center}
\includegraphics[scale=0.415]{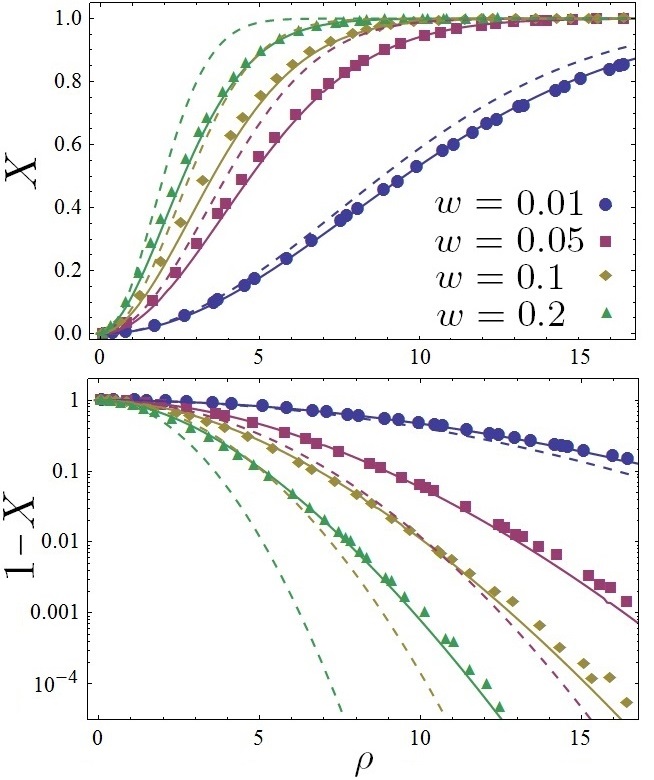}
\caption{\label{fig:numerics} Comparison of numerical integration of \eqref{X3} (solid curves), with computer simulations (solid markers), and the crude approximation of \eqref{XX} (dashed curves) for different values of $w=\{0.01, 0.05,0.1,0.2\}$ using $\eta=2$, $\beta=1$ and $\rho_{\A}= \rho_{\B}$ in a square domain as in Fig.~\ref{fig:setup}.}
\end{center}
\end{figure}

The simulation results are shown in Fig.~\ref{fig:numerics} using solid markers for different values of $w=\{0.01, 0.05,0.1,0.2\}$ using $\eta=2$, $\beta=1$ and $\rho_{\A}= \rho_{\B}$ in a square domain of side $L=10$ as in Fig.~\ref{fig:setup}.
Numerical integration of Eq.~\eqref{X3} is shown using solid curves, and is seen to be in very good agreement with the computer simulations thus verifying the semi-quenched nature of the keyhole system set-up.

The leading order prediction of Eq.~\eqref{XX} is shown in Fig.~\ref{fig:numerics} using dashed curves. 
As expected, this approximation is adequate at low densities $\rho$ and small openings $w$, but worsens when increasing either $\rho$ or $w$.
Indeed, if the distance $d_n$ of the nearest node to the opening is much larger than $w$ then $d_n\approx \rho^{-1/2}$ and there is little to no overlap between connection regions (see Fig.~\ref{fig:overlaps}.c.) and the independence assumption in Eq.~\eqref{X1} is reasonable.
To estimate the critical density $\rho_c$ at which \eqref{XX} begins to diverge we set $\rho_c=\rho_\A = \rho_\B$, and require $1-X=\mathcal{O}(1)$, (e.g. $X=e^{-1}$) giving $\rho_c \sim \sqrt{ \beta^{3/2} / w }$.
%, in good agreement with Fig.~\ref{fig:numerics}.

%%%%%%%%%%%%%%%%%%%%%%%%%%

\section{Conclusion}
\label{sec:concl}

In this paper we have investigated the connectivity properties of random networks in non-convex domains, and in particular when two networks in convex sub-domains may communicate through a small opening.
We have shown that the connectivity properties of the complete network is a product of three independent events, the connectivity of each sub-network and $X$, the probability that a communication link exists between them.
Furthermore, we have shown that whilst both sub-networks have nodes randomly distributed in space, connectivity across the complete network exhibits semi-quenched disorder as only nodes in one of the two sub-domains are frozen random variables during the link formation process. 
We have benefited through this reduction substantially, and have obtained general expressions for $X$ (see Eq.~\eqref{X3}) and confirmed them through computer simulations.

Further to the semi-quenched nature of the system, 
it seems clear that the keyhole problem is a challenging one due to the number of length scales involved.
We have the typical size of system which is about $\sqrt{V}$, the typical connectivity range $r_0=\beta^{-1/\eta}$, the typical distance between neighbouring nodes in two dimensions $\rho^{-1/2}$, and finally the width of the opening $w$.
Hence, asymptotic methods requiring significant separation between each of the scales makes further analysis a formidable challenge.
For example, even when $w \!\! \ll \!\! r_0$ there are \textit{a priori} three possible regimes: a) $\rho^{-1/2} \!\! \leq \!\! w$, b) $w \!< \! \rho^{-1/2} \! < \! r_0$, and c) $\rho^{-1/2} \geq r_0$.
Note that in physical systems, more length scales may become relevant, e.g. the signal wavelength.
Nevertheless, we have obtained leading order analytic expressions for $X$ (see Eq.~\eqref{XX}) valid up to densities $\rho_c \sim \sqrt{ \beta^{3/2} / w }$.

Familiar physically relevant settings where the above theory may be applicable are wireless networks in residential, business or industrial buildings. 
Here, scattered devices form a machine to machine communication network whose functionality can be disrupted by the topological features of the confining space, e.g. doorways and windows.
Perhaps a more suitable application derives from ad hoc networks with minimal configuration requirements and quick deployment rates, often used for emergency situations like natural disasters or military conflicts.
Here, the connectivity through small openings may correspond to that through narrow pathways, gaps and cracks encountered, e.g. in post earthquake urban environments, underground tunnels, or even inside the human body \cite{akyildiz2008nanonetworks}.
Indeed, modern swarms of small robots can communicate, sense, collaborate, fetch and carry \cite{nouyan2009teamwork}. 
These technological advancements lend themselves to search and rescue operations in hazardous environments \cite{cui2004swarm} and also to non-invasive medical diagnostics and treatments by so called ``nanobots" \cite{sajja2009development}.
On the theoretical side, we expect our results to lead to further insight and understanding on network connectivity in non-convex domains.

%%%%%%%%%%%%%%%%%%%%%%%%%%

\section*{Acknowledgements}
The authors would like to thank the directors of the Toshiba Telecommunications Research Laboratory for their support.

%%%%%%%%%%%%%%%%%%%%%%

\bibliographystyle{ieeetr}
\bibliography{mybib}

\begin{thebibliography}{10}

\bibitem{cordeiro2011ad}
C.~D.~M. Cordeiro and D.~P. Agrawal, {\em Ad hoc and sensor networks: theory
  and applications}.
\newblock World Scientific, 2011.

\bibitem{mohan2009extensive}
Y.~Mohan and S.~Ponnambalam, ``An extensive review of research in swarm
  robotics,'' in {\em Nature \& Biologically Inspired Computing, 2009. NaBIC
  2009. World Congress on}, pp.~140--145, IEEE, 2009.

\bibitem{li2009connectivity}
J.~Li, L.~Andrew, C.~Foh, M.~Zukerman, and H.~Chen, ``Connectivity, coverage
  and placement in wireless sensor networks,'' {\em Sensors}, vol.~9,
  pp.~7664--7693, 2009.

\bibitem{moslehi2010reliability}
K.~Moslehi and R.~Kumar, ``A reliability perspective of the smart grid,'' {\em
  Smart Grid, IEEE Transactions on}, vol.~1, no.~1, pp.~57--64, 2010.

\bibitem{harri2009mobility}
J.~Harri, F.~Filali, and C.~Bonnet, ``Mobility models for vehicular ad hoc
  networks: a survey and taxonomy,'' {\em Communications Surveys \& Tutorials,
  IEEE}, vol.~11, no.~4, pp.~19--41, 2009.

\bibitem{heidemann2012underwater}
J.~Heidemann, M.~Stojanovic, and M.~Zorzi, ``Underwater sensor networks:
  applications, advances and challenges,'' {\em Philosophical Transactions of
  the Royal Society A: Mathematical, Physical and Engineering Sciences},
  vol.~370, no.~1958, pp.~158--175, 2012.

\bibitem{penrose2003random}
M.~Penrose, {\em Random geometric graphs}, vol.~5.
\newblock Oxford University Press Oxford, UK:, 2003.

\bibitem{albert2002statistical}
R.~Albert and A.-L. Barab{\'a}si, ``Statistical mechanics of complex
  networks,'' {\em Reviews of modern physics}, vol.~74, no.~1, p.~47, 2002.

\bibitem{xue2004number}
F.~Xue and P.~R. Kumar, ``The number of neighbors needed for connectivity of
  wireless networks,'' {\em Wireless networks}, vol.~10, no.~2, pp.~169--181,
  2004.

\bibitem{coon2012impact}
J.~Coon, C.~Dettmann, and O.~Georgiou, ``Impact of boundaries on fully
  connected random geometric networks,'' {\em Physical Review E}, vol.~85,
  p.~011138, 2012.

\bibitem{coon2012full}
J.~Coon, C.~Dettmann, and O.~Georgiou, ``Full connectivity: corners, edges and
  faces,'' {\em Journal of Statistical Physics}, pp.~1--21, 2012.

\bibitem{haenggi2009stochastic}
M.~Haenggi, J.~Andrews, F.~Baccelli, O.~Dousse, and M.~Franceschetti,
  ``Stochastic geometry and random graphs for the analysis and design of
  wireless networks,'' {\em Selected Areas in Communications, IEEE Journal on},
  vol.~27, pp.~1029--1046, 2009.

\bibitem{keyholeorigin}
D.~Chizhik, G.~Foschini, and R.~Valenzuela, ``Capacities of multi-element
  transmit and receive antennas: Correlations and keyholes,'' {\em Electronics
  Letters}, vol.~36, no.~13, pp.~1099--1100, 2000.

\bibitem{nrayleigh}
J.~Salo, H.~El-Sallabi, and P.~Vainikainen, ``Statistical analysis of the
  multiple scattering radio channel,'' {\em Antennas and Propagation, IEEE
  Transactions on}, vol.~54, no.~11, pp.~3114--3124, 2006.

\bibitem{sornette2004critical}
D.~Sornette, {\em Critical phenomena in natural sciences: chaos, fractals,
  selforganization, and disorder: concepts and tools}.
\newblock Springer Verlag, 2004.

\bibitem{bocus2012keyhole}
M.~Z. Bocus, C.~P. Dettmann, J.~P. Coon, and M.~R. Rahman, ``Keyhole and
  reflection effects in network connectivity analysis,'' {\em arXiv preprint
  arXiv:1211.6255}, 2012.

\bibitem{tse2005fundamentals}
D.~Tse and P.~Viswanath, {\em Fundamentals of wireless communication}.
\newblock Cambridge University Press, 2005.

\bibitem{akyildiz2008nanonetworks}
I.~F. Akyildiz, F.~Brunetti, and C.~Bl{\'a}zquez, ``Nanonetworks: A new
  communication paradigm,'' {\em Computer Networks}, vol.~52, no.~12,
  pp.~2260--2279, 2008.

\bibitem{nouyan2009teamwork}
S.~Nouyan, R.~Gro{\ss}, M.~Bonani, F.~Mondada, and M.~Dorigo, ``Teamwork in
  self-organized robot colonies,'' {\em Evolutionary Computation, IEEE
  Transactions on}, vol.~13, no.~4, pp.~695--711, 2009.

\bibitem{cui2004swarm}
X.~Cui, T.~Hardin, R.~Ragade, and A.~Elmaghraby, ``A swarm-based fuzzy logic
  control mobile sensor network for hazardous contaminants localization,'' in
  {\em Mobile Ad-hoc and Sensor Systems, 2004 IEEE International Conference
  on}, pp.~194--203, IEEE, 2004.

\bibitem{sajja2009development}
H.~K. Sajja, M.~P. East, H.~Mao, A.~Y. Wang, S.~Nie, and L.~Yang, ``Development
  of multifunctional nanoparticles for targeted drug delivery and non-invasive
  imaging of therapeutic effect,'' {\em Current drug discovery technologies},
  vol.~6, no.~1, p.~43, 2009.

\end{thebibliography}

\end{document}